\documentclass[12pt,preprint]{aastex}

\slugcomment{Accepted to ApJ} 

\shorttitle{Starspots Detected During Consecutive Transits}
\shortauthors{Dittmann et al..}

\begin{document}

\title{A Tentative Detection of a Starspot During Consecutive Transits of an Extrasolar Planet from the Ground: No Evidence of a Double Transiting Planet System Around TrES-1}

\author{Jason A. Dittmann\altaffilmark{1}, Laird M. Close\altaffilmark{1}, Elizabeth M. Green\altaffilmark{1}, Mike Fenwick\altaffilmark{1}}

\email{dittmann@email.arizona.edu}

\altaffiltext{1}{Steward Observatory, University of Arizona, Tucson, AZ 85721}

\begin{abstract}
There have been numerous reports of anomalies during transits of the planet TrES-1b. Recently, Rabus and coworkers' analysis of HST observations lead them to claim brightening anomalies during transit might be caused by either a second transiting planet or a cool starspot. Observations of two consecutive transits are presented here from the University of Arizona's 61-inch Kuiper Telescope on May 12 and May 15, 2008 UT.  A $5.4 \pm 1.7$ mmag ($0.54 \pm 0.17 \%$) brightening anomaly was detected during the first half of the transit on May 12 and again in the second half of the transit on May 15th. We argue that the significance of these spot events are 3.2 and 2.9 $\sigma$ for May 12 and May 15 respectively and we estimate that each of these have a probability $\geq 90\%$ of not being systematic red noise peaks. Therefore we conclude that this is a tentative detection of a r $\geq 6 R_{\earth}$ starspot rotating on the surface of the star. We suggest that all evidence to date suggest TrES-1 has a spotty surface and there is no need to introduce a second transiting planet in this system to explain these anomalies. Assuming that the spin axis of the star and orbital axis of the planet are aligned (and in the plane of the sky) suggests a stellar rotational period of $40.2 \pm 0.1$ days. Introducing the $\lambda = 30 \pm 21^{\circ}$ inclination of the stellar spin axis with respect to the planetary orbital axis of Narita et al. (2007) adds much more uncertainty and we are only able to constrain the rotational period of the star to $40.2^{+22.9}_{-14.6}$ days, which is consistent with the previously observed $P_{obs} = 33.2^{+22.3}_{-14.3}$ day period. We note that this technique could be applied to other transiting systems for which starspots exist on the star in the transit path of the planet in order to constrain the rotation rate of the star.

\end{abstract}


\section {Introduction}

TrES-1b is a Jupiter sized extrasolar planet orbiting a 12th magnitude K0 V star, discovered by Alonso et al. (2004) using the transit search method. Since then, there have been numerous investigations regarding photometric anomalies during transits of TrES-1b. HST ACS spectra analyzed in Rabus et al. (2009) show a very significant ($\sim 2.7 \pm 0.2$mmag) brightening anomaly just before the center of transit. Professional and amateur data taken and compiled by Price et al. (2006) show a $\sim 5$ mmag brightening event during the egress of the transit. Studying of anomalies just outside of egress led to the suggestion of trojan-like objects, which have been constrained by Ford and Gaudi (2006). In order to understand the strange anomalies inside the transit light curve, Winn et al. (2007) obtained transit light curves in the far red z-band for three consecutive transits of TrES-1b with the FWLO 1.2 m and KeplerCam. Their light curves had low scatter, $\sim 1$ mmag, and they found no evidence for the existence of anomalies either inside or outside of the transit over a 9 day period. However, if those anomalies were caused by the planet transiting in front of a starspot, then observations in the z-filter used by Winn et al. (2007) would be much less sensitive to starspots than the bluer filters used in the previous studies (or there were no spots visible over those 9 days). Rabus et al. (2009), show in HST ACS spectral data that their anomaly is sensitive to wavelength, with a peak in the H$_{\alpha}$ bin (as expected for a cool starspot, and as demonstrated in a similar study by Pont et al. (2007)), though they argue this peak might not be real due to higher noise in that wavelength bin. However, they do find that the brightening event is smaller for longer wavelengths (Rabus et al. (2009)), suggesting that Winn et al. (2007)'s z-band data may not have been sensitive to such spots on TrES-1.

Rabus et al. (2009) present two possible explanations for these anomalies; starspots, or the first double transiting planetary system. They argue that a starspot crossing event would be wavelength dependent while a double transit would not. Also, starspots on the star would appear in different locations from transit to transit, while a planet-planet occultation would be a very rare event unless the planets were in a resonant orbit. Rabus et al. (2009) also modeled a double planet transit to simulate their data and were able to create a lower bound on the radius of the second planet of 1.081 $R_{J}$. They argue that more observations are required to definitively understand the nature of the HST anomaly.

Our own group observed a very similar anomaly to that of Rabus et al. (2009) in April 2007 in the R band. This observation prompted follow-up observations. In this paper we present evidence that these brightening anomalies are likely spots on the star.

By observing a starspot group on the surface of the star as a transiting planet occults them on multiple consecutive transits, it is possible, in theory, to measure this rotational period (as recently outlined in Silva-Valio (2008)). Studies measuring the Rossiter-McLaughlin effect (Winn 2006) measure the tilt of the stellar rotational axis with respect to the planetary orbital axis in the plane of the sky. This effect was measured by Narita et al. (2007) for TrES-1 who found $\lambda = 30 \pm 21^{\circ}$ (or $24 \pm 23^{\circ}$ dropping one outlier). The orbital axis of TrES-1b with respect to the line of sight is $88.5^{\circ +1.5}_{-2.2}$ (Alonso et al. 2004). Laughlin et al. (2005) found V$_{*}sin(i_{*}) = 1.08 \pm 0.3$ km/s.  

We have observed 2 consecutive transits of TrES-1b, and are able to conclude that the increased flux anomalies during consecutive transits of TrES-1b are likely due to the planet crossing in front of a starspot rotating on the star surface and not due to a double planet transit.

\section{OBSERVATIONS \& REDUCTIONS}

Data was taken at the University of Arizona's 61-inch Kuiper telescope on Mt. Bigelow on April 14, 2007, May 12, 2008 and May 15, 2008 UT with the Mont4k CCD, binned to 0.43"/pix, in the R band. There were light clouds on May 12 ($<$ 1 magnitude extinction) but photometric conditions on April 14, 2007 and May 15, 2008. 

The images were bias subtracted, flat-fielded, and bad pixel cleaned in the usual manner. Photometry (and photometric errors) analysis was executed on all images using the DAOPHOT package in IRAF. Relative photometry was done using reference stars in the field of view. The reference stars were normalized to unity and then weighted according to their average fluxes.

We applied 2$\sigma$ clipping to the reference stars. However, data points for TrES-1 were not clipped. The final light curve for TrES-1 was normalized by division of the weighted average of four reference stars. On May 12 only two of the four May 15 reference stars were used since, due to the patchiness of clouds, we utilized the reference stars that were closest to TrES-1.

The light curves in Figure \ref{transits} have a photometric RMS range of $\sim1-2$ mmag and a time sampling of $\sim40$ seconds. This is very typical of the relative photometric precision achieved with the Mont4k on the 61" telescope for high S/N images (Randall et al. (2007), Dittmann et al. (2009)).

\section{ANALYSIS}
The planetary transit light curves were fit using the $\chi^{2}$ method prescribed by Mandel and Agol (2002). The transit TrES-1b parameters used in the fit were those compiled by Butler et al. (2006) from Alonso et al. (2004), and the impact parameter, $0.18 \pm 0.1$, was provided by Gaudi and Winn (2007). Linear and quadratic limb darkening parameters were taken from Claret (2000). In order to understand and characterize departures from the transit related to starspot occultation, the only parameter that was allowed to vary in the fit was the central time of each transit, $T_{c}$. The time of the center of each of these transits ($T_{c}$), are shown in Table \ref{tc}. We note that the purpose of this paper was not to re-derive the parameters of the transit but to understand if these anomalies are real.

\begin{figure}
\epsscale{1.0}
\plotone{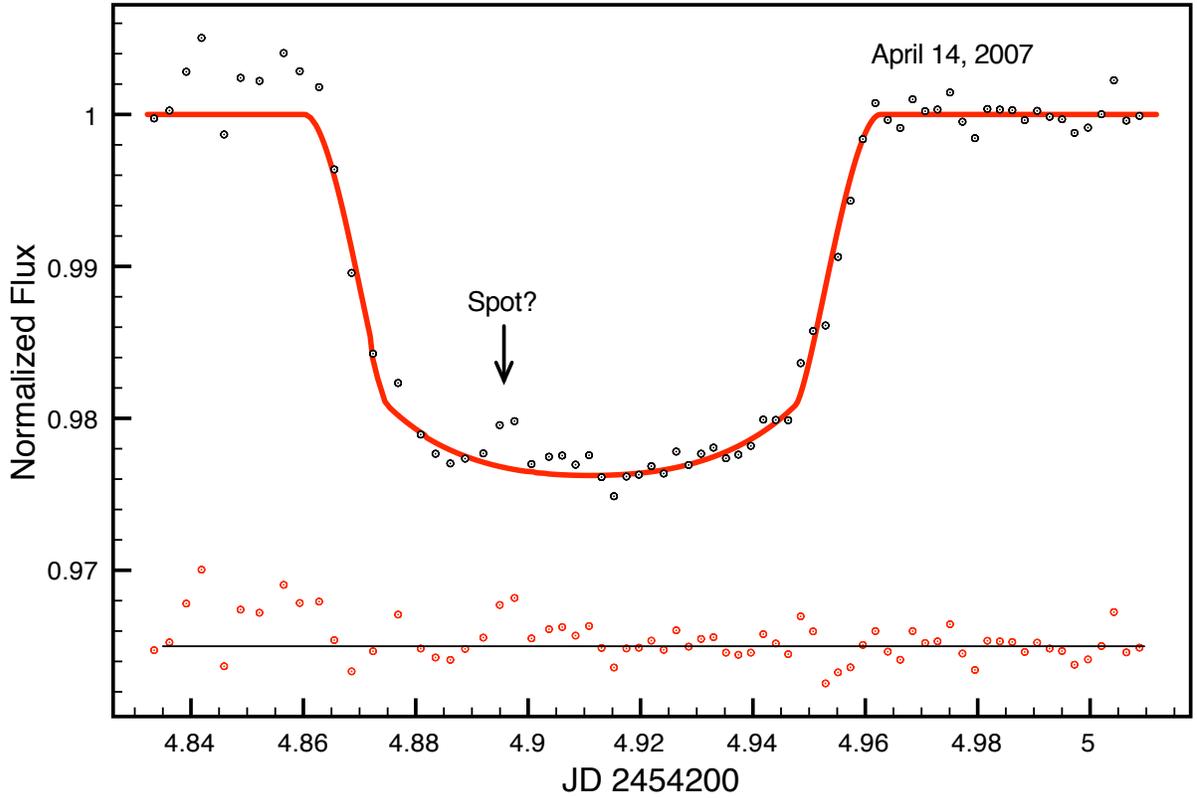}
\caption{
Transit light curve of TrES1-b taken on April 14, 2007 UT at the University of Arizona's 61 inch Kuiper telescope in the R filter. Residuals of the curve from the fit are shown below. There is a tentative, $\sim 2.8$ mmag, brightening event just before the center of the transit. This prompted the follow-up observations of consecutive transits in 2008.
}
\label{Fenwick}
\end{figure}

\begin{figure}
 \epsscale{1.0}
 \plottwo{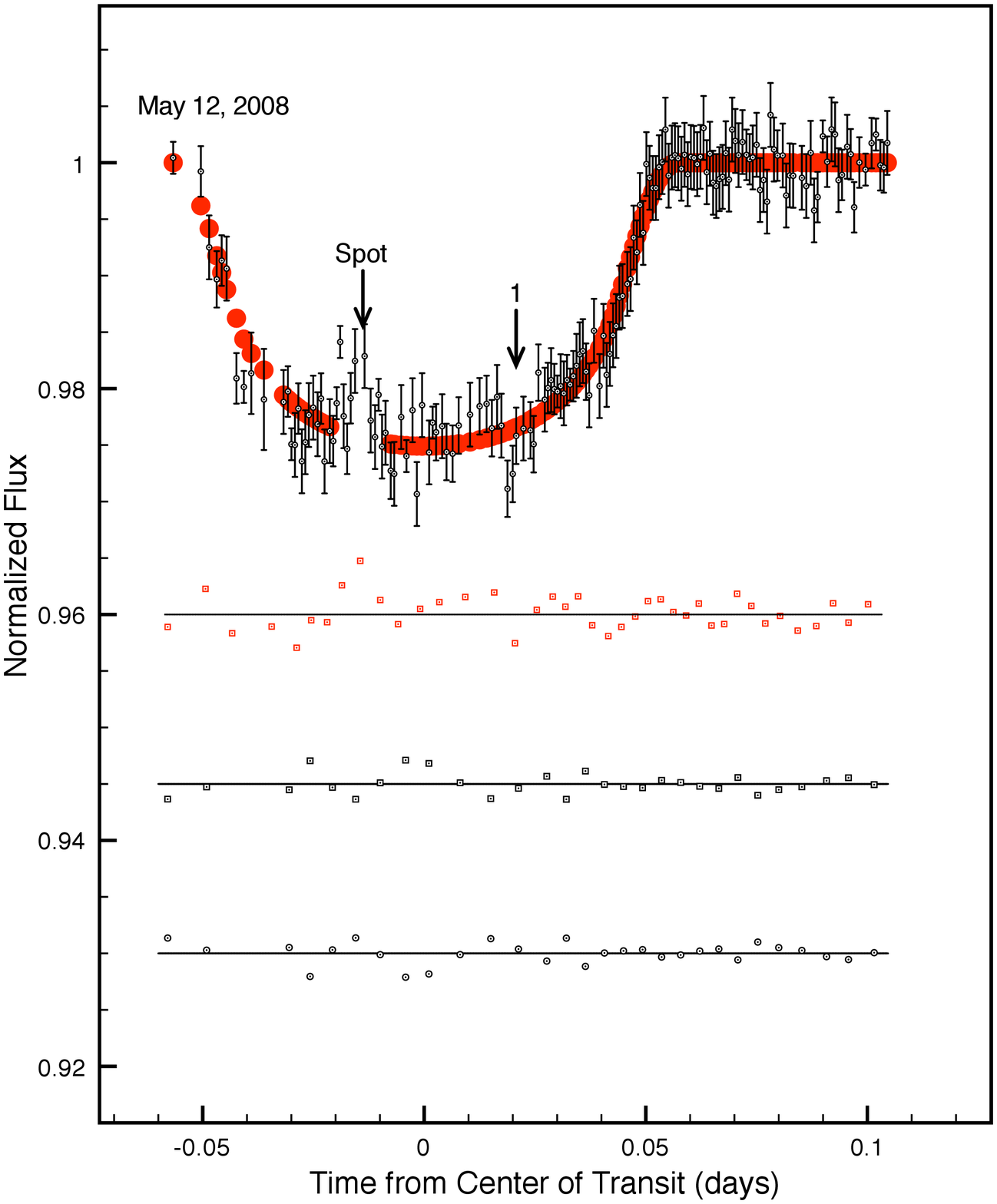}{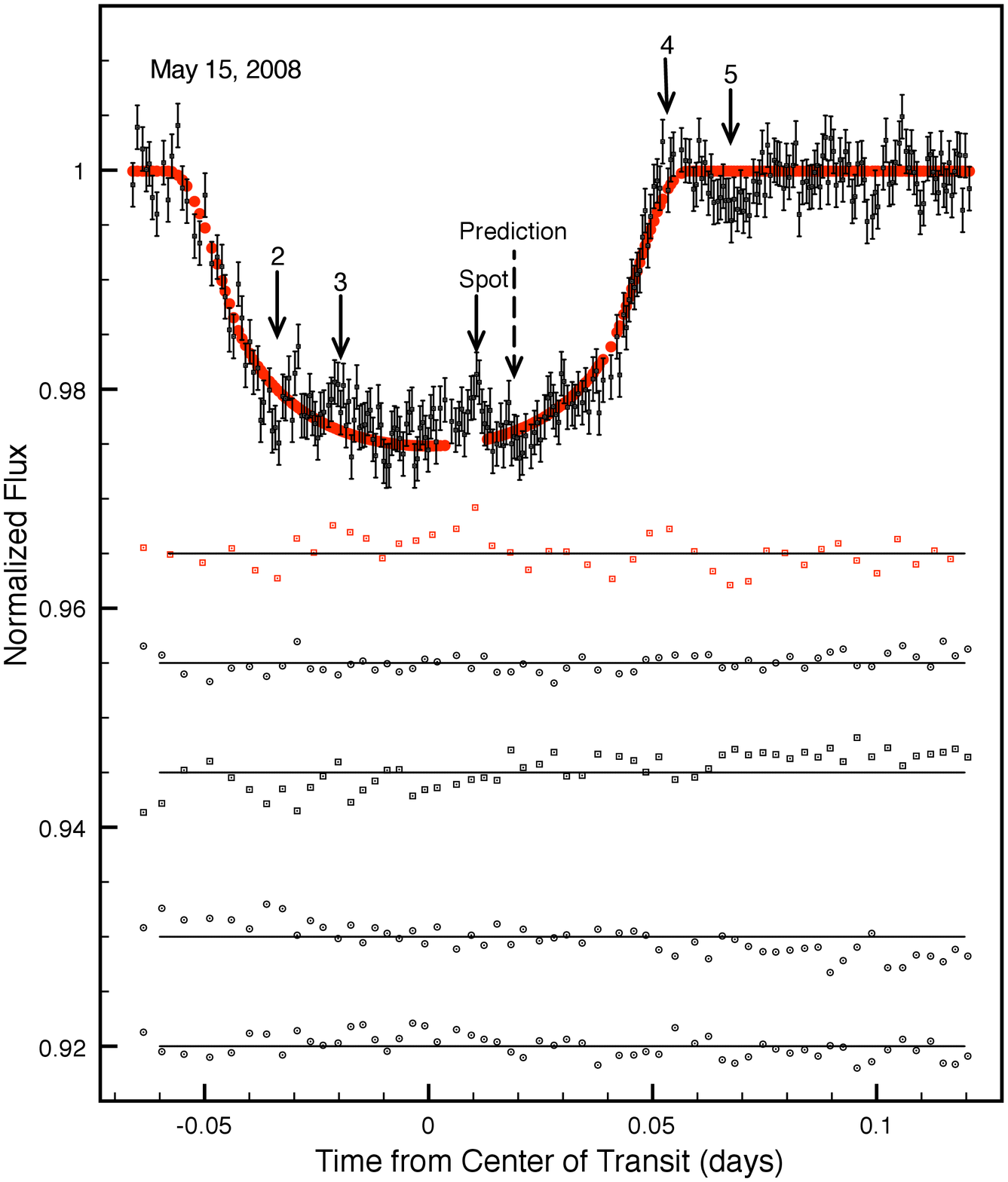} 
 \caption{
Data collected with the Mont4k CCD at the University of Arizona's 61 inch Kuiper telescope in the R filter. The left curve is data collected on May 12, 2008 and the right curve is from May 15, 2008. Beneath each curve are the red data points of 5 minute discrete binned residuals of the fit. We note that we used higher (15 minute) binning just at ingress of the May 12 night due to the higher air-mass at the beginning of that observation. The anomaly is detected as a 3.2$\sigma$ event on May 12 and a 2.9$\sigma$ event on May 15 (see Figure \ref{gauss}). The predicted position of the spot on May 15 is just 27 minutes later (well within the error of $P_{obs}$) than actually observed (assuming the observed period of $33.2^{+22.3}_{-14.3}$ days with $I = \lambda = 30^{\circ}$, $V_{*} = 0.935$ km/s, (Narita et al. (2007), Laughlin et al. (2005))). The black points beneath each transit curve represent the reference star photometry measured for each transit.
 }
 \label{transits}
 \end{figure}

On April 14, 2007 UT, a transit of TrES-1b was observed at the University of Arizona's 61-inch Kuiper telescope. The data showed a brightening just before the center of the transit similar to that in Rabus et al. (2009) (see Figure \ref{Fenwick}). We found the peak of the brightening to be $\sim 2.3$ mmag. Since this was similar in amplitude and duration to the HST anomaly of Rabus et al. (2009) this motivated follow-up observations during May 2008. In particular, if these anomalies were due to a starspot (as first suggested by Charbonneau et al. (2006) to explain the HST event), then the anomaly may appear later in phase in consecutive transits. We will focus on the pair of consecutive transits observed in 2008 for the rest of this paper.

\subsection{Could These Consecutive Anomalies Be Due to a Starspot Rotating on the Surface of TrES-1?}
There is a $5.4 \pm 1.7$ mmag brightening anomaly during the first half of the transit on May 12, and one in the second half of the transit on May 15. Assuming that these events are the planet passing in front of a darker part of the star (likely a starspot), and that the starspot seen on May 15 is the same spot as that first seen on May 12, and that there is no significant spot evolution in the 3 days between transits, then it is possible to estimate the rotation rate of the star TrES-1 if we know or can estimate the relative inclination, $\lambda$, of the star's spin axis and the planet's rotational axis. However, these are difficult anomalies to measure and we estimate the likelihood of these events being noise in section 4.2. We assume for the rest of this section that they are real. Also, for the rest of this paper we will make the approximation that the planetary orbital axis is $90^{\circ}$ inclined from the line of sight (instead of the measured $88.5^{\circ +1.5}_{2.2}$ by Alonso et al. (2004)). Because the errors in the the location of the stellar rotation axis are large, this approximation will have little effect on our final result. Also, without knowing the starspot geometry, these formulas are simple approximations. We will use the following notation (see Figure \ref{angles}): 

\begin{enumerate}
\item $I$ is the inclination of the stellar spin axis out of the plane of the sky ($i_{*} = 90 - I$).
\item $\lambda$ is the sky-projected angle between the stellar spin axis and the planetary orbital axis, where we have made the good approximation that the planetary orbital axis is in the plane of the sky.
\end{enumerate}

\subsubsection{The $\lambda = 0^{\circ}$, $I = 0^{\circ}$, $i_{*} = 90^{\circ}$ Case}
We first examine the simpler case where the axes are aligned, and that the inclination of both axes along the line of site is effectively $90^{\circ}$ (in the plane of the sky).

\begin{figure}
\epsscale{1.0}
\plottwo{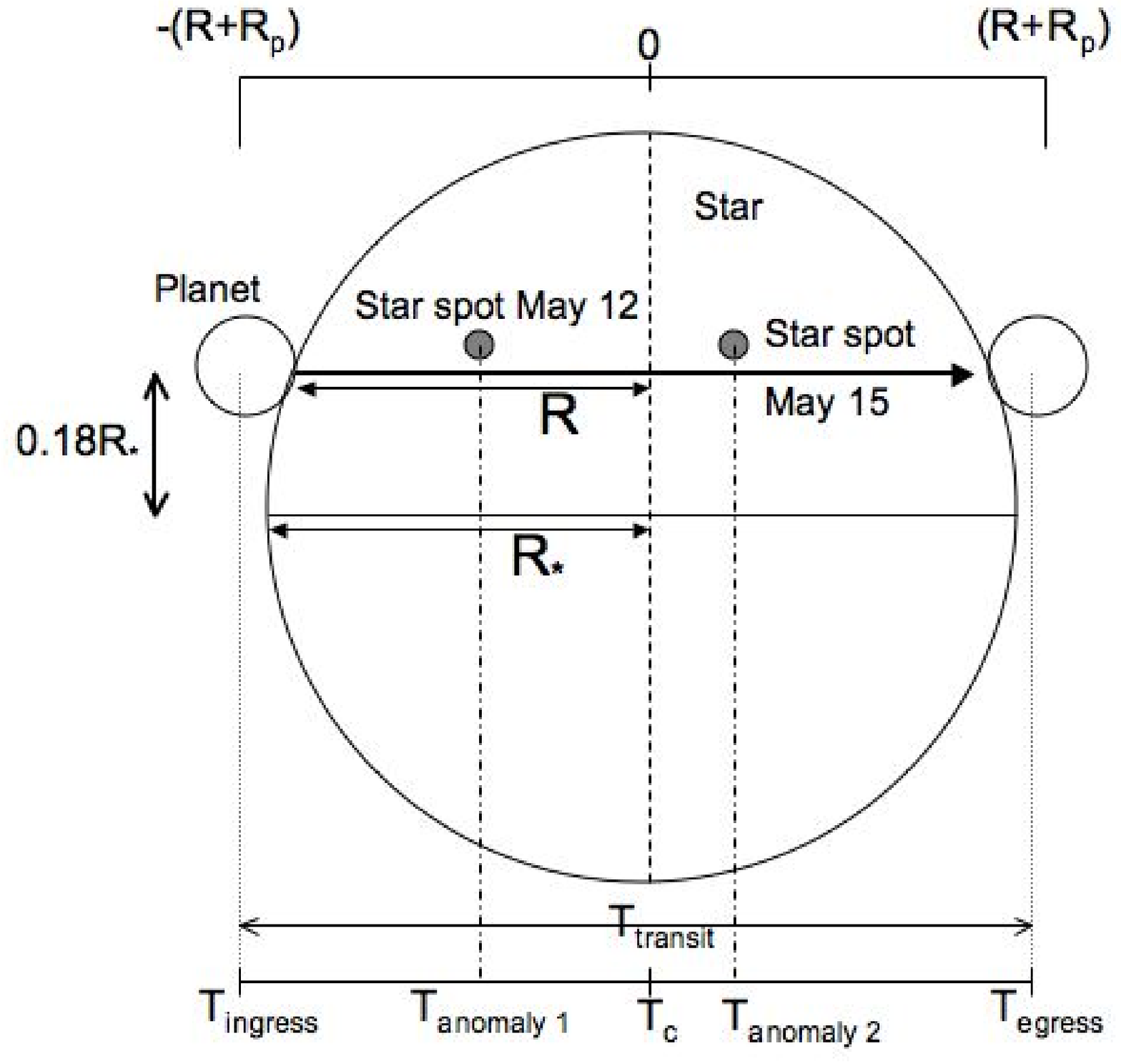}{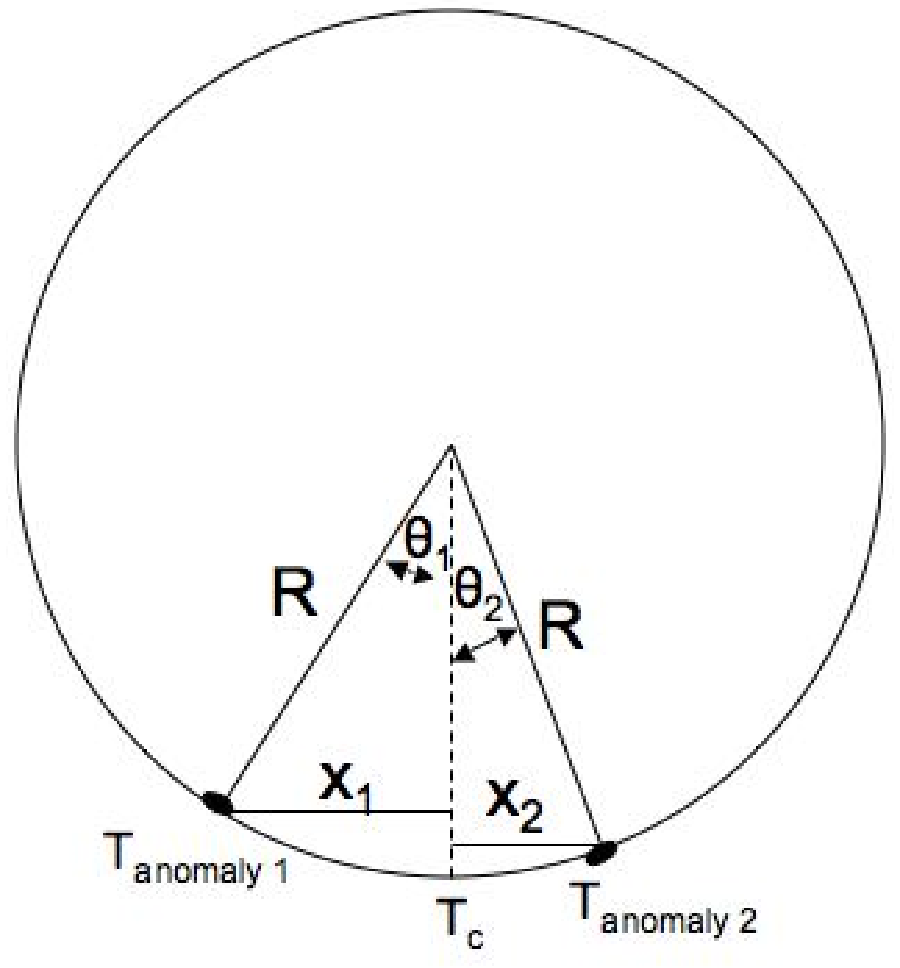}
\caption{
\textbf{Left:} Geometry of the transit ($\lambda = 0$, $I = 0$ case). The planet moves across the face of the star with an impact parameter $b = 0.18$. During each transit, it passes in front of a starspot on the face of the star. By comparing the location of the brightening event associated with the star spot during each transit, we can calculate the rotational period of the star at that latitude. Since the star is likely at an inclination of $\sim 88.4^{\circ}$, we assume it is exactly edge-on ($90^{\circ}$). 
\textbf{Right:} Star spot location during each transit is characterized by an angle, $\theta$. Perspective here is looking down the rotational axis of the star. $R=0.984R_{*}$.
}
\label{fig1}
\end{figure}

The geometry of this case is illustrated in Figure \ref{fig1}. Here, R is the radius of the circular crossection at the transit latitude and is given by $R = R_{*}cos($lat$)$, where the latitude is calculated from the impact parameter. For TrES-1, this yields $R = 0.984 R_{*}$. The time of the peak of the anomaly during the transit is directly related to the projected location of the spot on the surface of the star, which can then be used to calculate the angle from the apparent central line of longitude from Earth. This is shown in figure \ref{fig1}. If we assume that the planet's orbital axis and TrES-1's spin axis are perfectly aligned in the plane of the sky ($\lambda = 0 \pm 0^\circ$), we can calculate the rotational period of the star at that latitude as:

\begin{equation}
P_{rot} = \frac{2\pi(T_{anomaly_{2}} - T_{anomaly_{1}})}{\theta_{2} + \theta_{1}} \mbox{           Case: $\lambda = 0$, $I = 0$}
\label{Period1}
\end{equation}

where:

\begin{equation}
\theta_{i} = sin^{-1} \left( 2 \left[ 1+\frac{R_{p}}{R_{*}cos(lat)} \right] \left[ \frac{|T_{anomaly,i}-T_{c}|}{T_{transit}} \right] \right) \mbox{ rad}
\label{theta}
\end{equation}

Here, the factor of $\frac{R_{p}}{R_{*}cos(lat)}$ corrects for the fact that ingress is when the leading limb of the planet crosses the star and egress is when the trailing limb of the planet leaves the star. In other words, the distance traveled in the transit time ($T_{transit}$ = 0.115 days or 2.76 hours for TrES-1b) is the time required to move $2R_{p} + 2R_{*}cos(lat)$ (see Figure \ref{fig1}).

After fitting with the method of Mandel and Agol (2002), we obtained an initial reduced $\chi^{2}$ of 0.88 for May 12 and 0.97 for May 15. The $\chi^{2}$ was minimized, and $T_{c}$ measured, excluding the starspot anomaly points for the fit. See Table 1 for the determined transit center $T_{c}$ values. 

In order to characterize the range of stellar rotation rates that we are sensitive to, we find the maximum rotation rate that TrES-1 could have in order for the starspot seen on the first night to completely rotate off the visible side of the star for the observations on May 15. In other words, $\theta_{2} = 90^{\circ}$, and the rotational axis is in the plane of the sky so that the spot moves the maximum distance between transits. We find that the star would need to be rotating once every 10.03 days in order for this to occur. Consequently, we are then sensitive to periods $> 10$ days. 

Using equations \ref{Period1} and \ref{theta}, which assume that the rotational axis of the star is in the plane of the sky, we find a rotation rate for TrES-1 of $40.2 \pm 0.1$ days ($\lambda = 0 \pm 0^{\circ}$). Alternatively, it is possible that this star rotated through any number of whole rotations plus the observed angle change observed. If the star had completed one additional rotation between the times of observation, this would correspond to a rotation rate of 2.84 days, which we consider to be very unlikely for an old K0V star and is inconsistent with other periods measured for this system.

\subsubsection{The $\lambda = 30 \pm 21^{\circ}$, $I = 0^{\circ}$, $i_{*}=90^{\circ}$ Case}
It has been estimated by Narita et al. (2007) that the rotational axis and the orbital axis of the planet TrES-1b are possible misaligned. They have estimated that the sky projected angle between the orbital axis of TrES-1b and the spin axis of TrES-1, $\lambda$ is $30 \pm 21^{\circ}$ (see Figure \ref{angles}). Now our time stamp measurements of each starspot occultation lie along the transit path while the starspot itself travels a shorter path. Therefore, the true longitudinal change differs from the $\lambda = 0$, $I=0$ case by a factor of $\sim$cos($\lambda$). Replacing $\theta_2+\theta_1$ in equation \ref{Period1} with $(\theta_2+\theta_1)cos(\lambda)$ yields:

\begin{equation}
P_{rot} \approx \frac{2\pi(T_{anomaly_{2}} - T_{anomaly_{1}})}{(\theta_{2} + \theta_{1})cos(\lambda)} \mbox{        $\lambda = 30 \pm 21^{\circ}$, $I = 0$}
\end{equation}

Using the values for $\lambda$ from Narita et al. (2007), we find a period of $46.4^{+23.6}_{-0.5}$ days. Since in this case, $I = 0$, Laughlin et al. (2005)'s value of V$_{*}sin(i_{*}) = 1.08 \pm 0.3$ km/s becomes V$_{*}$ = $1.08 \pm 0.3$ km/s because the rotational axis is in the plane of the sky (or $i_{*} = 90^{\circ}$). This yields an observed period of the star of $P_{obs} = 38.4^{+12.7}_{-8.4}$ days. Our value is slower, but still consistent with this observed period.

\subsubsection{The $\lambda = 0^{\circ}$, $I =30 \pm 21^{\circ}$, $i_{*}=60 \pm 21^{\circ}$ Case}
We now move on to the case where $\lambda = 0$ but the inclination of the stellar rotation axis out of the plane of the sky is nonzero. While we have no prior constraints on this angle, $I$, we will make the assumption that it is the same order as Narita et al. (2007)'s measurement of $\lambda$, namely $30 \pm 21^{\circ}$.

In the non-zero $I$ case, the rotational axis of the star is at a projected vertical position of $R_{*}cos(I)$. When connected with the positions of the starspots, this forms a triangle, where the subtended angle is now larger than for the $I = 0$ approximation. The $x_i$ value can still be taken from the time-stamp of the occultation by the planet (see eqn \ref{theta}), but the longitudinal angle of the starspot increases so $\theta_2+\theta_1 \sim (\theta_2+\theta_1)/cos(I)$. Then, the period becomes

\begin{equation}
P_{rot} \approx \frac{2 \pi cos(I) (T_{anomaly_{2}} - T_{anomaly_{1}})}{\theta_2+\theta_1} \mbox{       $\lambda = 0$, $I = 30 \pm 21^{\circ}$}
\label{newperiod}
\end{equation}

Assuming $I = 30 \pm 21^{\circ}$ and equation \ref{newperiod} we find a rotational period of $P_{rot} \approx 34.8^{+4.9}_{-9.5}$ days. Using this inclination for the Laughlin et al. (2005) value of V$_{*}sin(i_{*}) = 1.08 \pm 0.3$ km/s yields a rotation rate of $P_{obs} = 33.2^{+22.3}_{-14.3}$ days, taking into account the error in $i = 90 - I$ and the error in the measurement itself. Our measured $P_{rot}$ value is also consistent with this observed $P_{obs}$ value.

\subsubsection{The $\lambda = 30 \pm 21^{\circ}$, $I = 30 \pm 21^{\circ}$ Case}
In general, both $\lambda$ and $I$ are nonzero. Increasing $\lambda$ tends to result in a lengthening of the period while Increasing $I$ tends to lower the period. These two effects tend to cancel each other out if $I \sim \lambda$ and combining them yields:

\begin{equation}
P_{rot} \approx \frac{2 \pi cos(I) (T_{anomaly_{2}} - T_{anomaly_{1}})}{(\theta_2+\theta_1)cos(\lambda)} \mbox{       $\lambda = 30 \pm 21^{\circ}$, $I = 30 \pm 21^{\circ}$}
\label{finalperiod}
\end{equation}

Using this, we find $P_{rot} \approx 40.2^{+22.9}_{-14.6}$ days with large uncertainties. We will quote this result for the rest of this paper.

\subsection{Could These Consecutive Anomalies Be Due to a Double Transiting Planet?}
Rabus et al. (2009) postulated that the in-transit anomaly of TrES-1b they observed with HST could be due to another transiting planet in the system, and the brightening effect due to TrES-1b partially passing behind the secondary planet in our line of sight. Assuming this ''double eclipse'' theory is correct and that both of our events were also caused by such double eclipses requires that the outer planet, "TrES-1c" in 3.05 days covers just 0.464$R_{*}$ laterally, while TrES-1b only requires 0.115 days to cover 2.26$R_{*}$. This suggests that TrES-1c only has an orbital velocity of $\sim 1.1$ km/s and a semi major axis of $\sim 745$ AU and an orbital period of $\sim 21000$ years. It is clearly impossible to have such a system also produce the brightening event observed by Rabus et al. (2009) in 2004. This means that if this was a double transiting system, the TrES-1c orbital velocity required to explain the pair of 2008 events would be too slow to explain the appearance of the prior HST brightening event prior. 

It is also possible that one of our events could be due to "TrES-1c" while the other due to a starspot occultation. However, to have these two unrelated events happen in consecutive transits at a separation completely consistent with the rotation rate of the star is very unlikely. Therefore, we find that the most likely explanation for all of the anomaly observations in Table \ref{table} is that TrES-1b often occults a large starspot(s) on the surface of TrES-1.

\section{DISCUSSION}
\subsection{Rotational Period of a Starspot}
We assumed in calculating the $P_{rot} \approx 40.2^{+22.9}_{-14.6}$ day rotational period of the star that there was no significant spot evolution between observations. Starspot migration we consider to be irrelevant between observations since only a few tenths of a degree at most could be expected (Rodono et al. (1995), Strassmeier et al. (2003)).  Even in the event of the starspot slightly evolving during those 3.05 days, it would only introduce small errors relative to the errors introduced due to the large uncertainty in $\lambda$ into our calculations.

By approximating the starspot as 100\% black, we can estimate the minimum size of the spot. The peak brightening was $5.4$ mmag, and by dividing this by the depth of the main transit, $\sim 25$ mmag,  we are able to estimate the size of the spot in terms of the size of the planet. We find that the spot is $\geq 6 R_{\earth}$ in radius. This is a large spot for our Sun, but could be common for TrES-1. We note that the HST observations were best fit by a similar sized spot of $6.6 R_{\earth}$ (Rabus et al. (2009). 

We consider it unlikely that the brightening anomalies associated with TrES-1b transits are due to a double transiting planet as suggested by Rabus et al. (2009) because of the detection of a similar anomaly showing up during two consecutive transits that we associate. During the 3.05 days between the two anomaly observations, "TrES-1c" would have moved off the face of the star from the line of sight and TrES-1b would not have passed partially behind it during the May 15 transit. Therefore, a second brightening event would have been absent in the next period of the inner planet. However, we see a brightening event of the same size and duration in consecutive transits moving with the rotational period of the star. This is only consistent with the effect of a starspot rotating on the star, and not with a double planet occultation. Of course, we cannot say this is the same spot observed in 2004, but we can say that the surface of TrES-1b quite often contains at least one large starspot and so there is no need for a second transiting planet.

\subsection{Could this Pair of Anomalies be Random or Systematic Noise?}
The duration of the brightening event associated with the occultation of a starspot is dependent upon the relative geometry of the transit and the spot. The duration is equal to the planet crossing time across the size of the spot projected into the plane of the sky, while the amplitude of the event is dependent upon the temperature of the spot and the fraction of the sky projected starspot that is occulted by the planet. As the occultation becomes more grazing the amplitude of brightening event tends towards zero. We can estimate the duration of an occultation as the time it would take for TrES-1b to cross itself  as $\sim \frac{R_{p}}{v_{p}} \sim \frac{R_{p}}{R_{*}}T_{transit} \sim 22$ minutes. The duration of the starspot events on May 12 and May 15 are $\sim 20$ minutes, which is a reasonable duration for a $5.4$ mmag event.

While the spot occultation duration argument above gives some comfort, we must estimate the probability that these events are real. Pont and Moutou (2007) state that aperture photometry can give accurate results if they are de-correlated with atmospheric variations. Plots of the FWHM of the TrES-1 PSF vs. the unbinned residuals from the transit fit are shown in Figure \ref{fwhm}. The correlation coefficient from Neter et al. (1982) for all the points is -0.25 and 0.022 for May 12 and May 15 respectively. For only the anomaly (red) data points, the coefficients are -0.212 and -0.049. Therefore, from Figure \ref{fwhm} we conclude that systematic effects associated with the atmosphere or PSF are not creating these anomalies. We note that while other systematic noise due to other effects such as absolute sub-pixel position are not examined here, we find them likely relatively unimportant.

\begin{figure}
\epsscale{1.0}
\plottwo{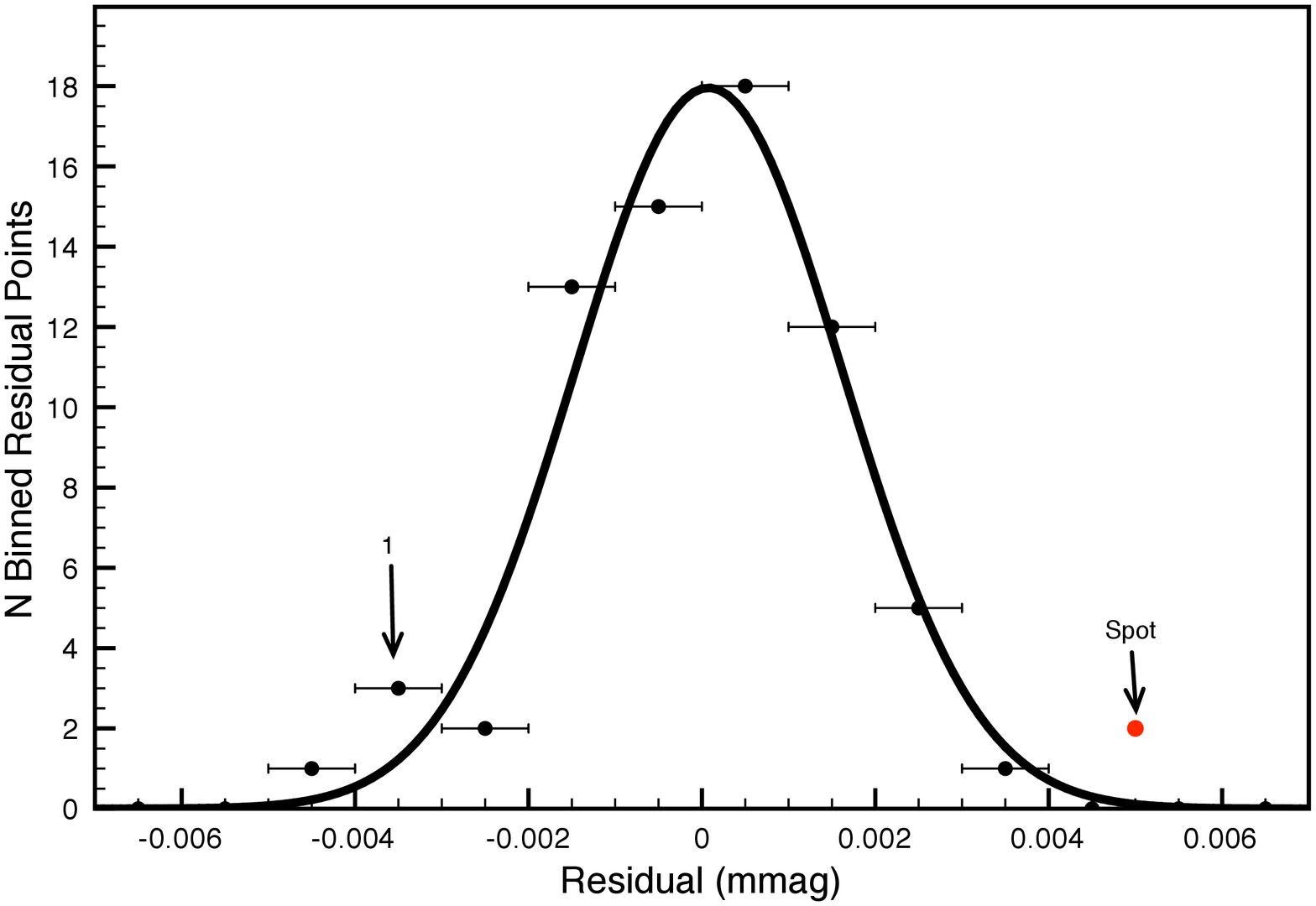}{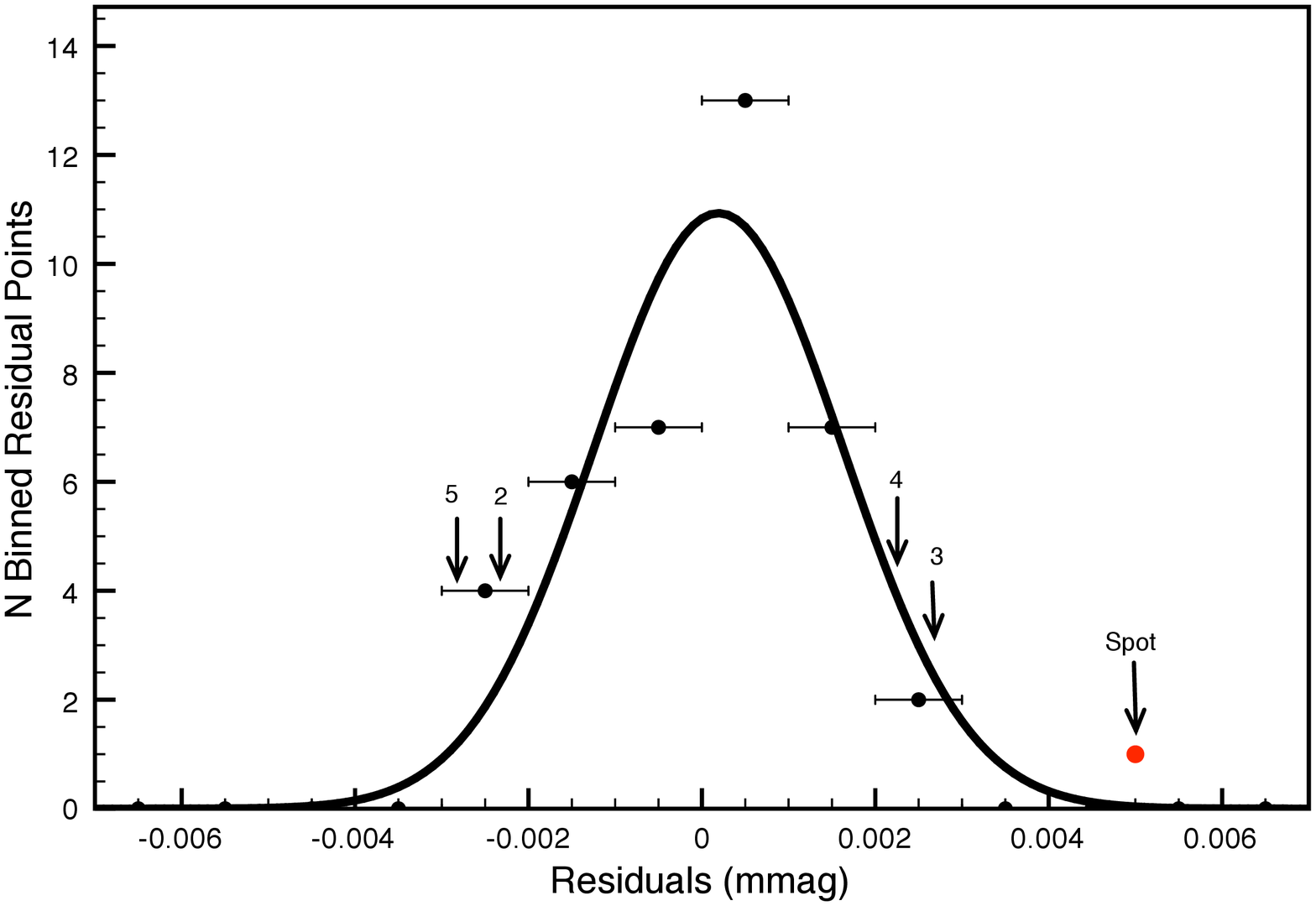}
\caption{
Gaussian fits of the binned residuals for each transit (including the spot events), May 12 on the left and May 15 on the right. These include only the residuals after ingress on May 12. The line represents the Gaussian fit, where the black data points are the residuals in each interval. The red points (at $3.2 \sigma$ and $2.9 \sigma$) represent the peak of the binned spot anomaly data points. The black numbers represent the location of other departures from transit which we associate with red noise peaks (see Figure \ref{transits}).}
\label{gauss}
\end{figure}

\begin{figure}
\epsscale{1.0}
\plottwo{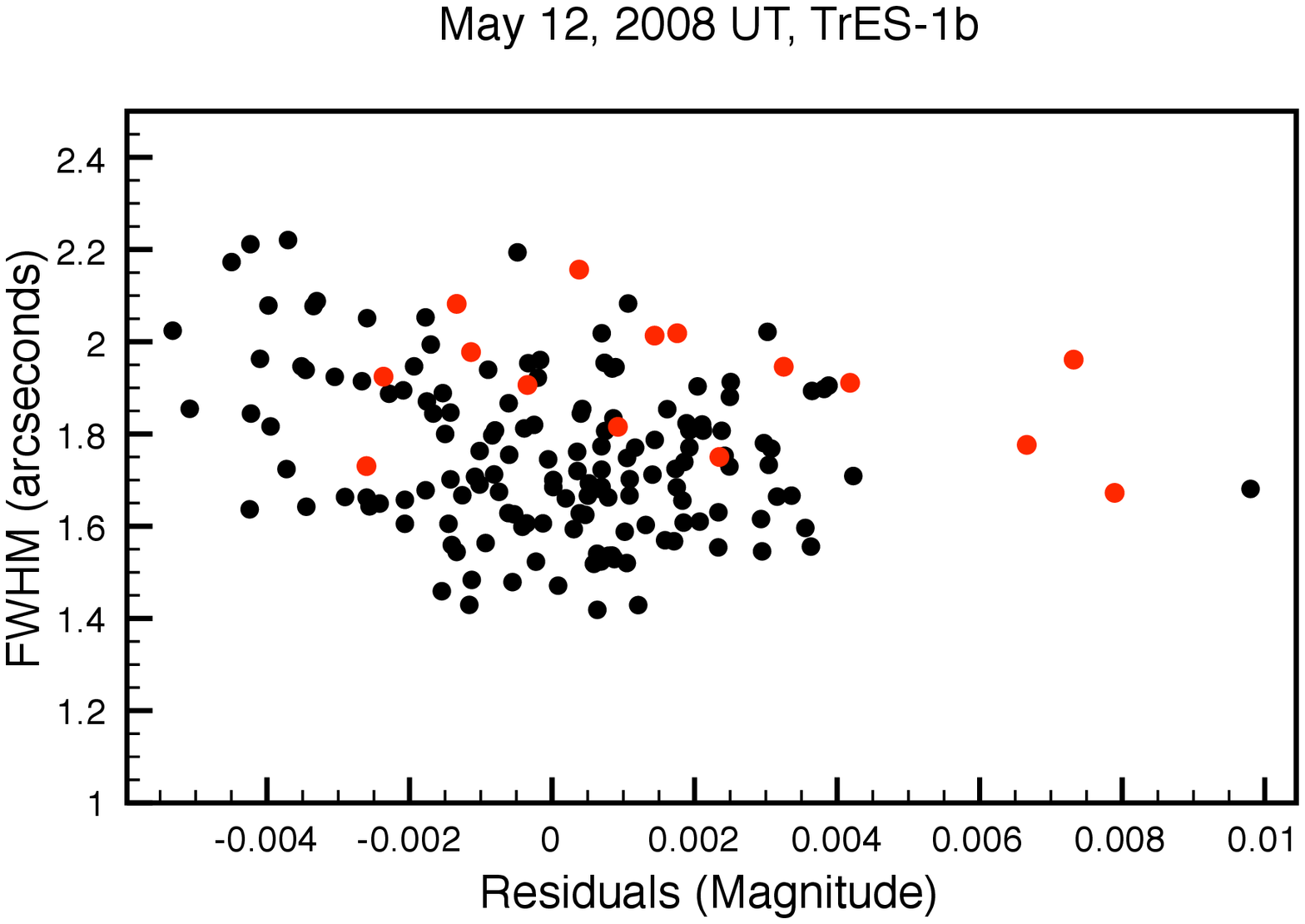}{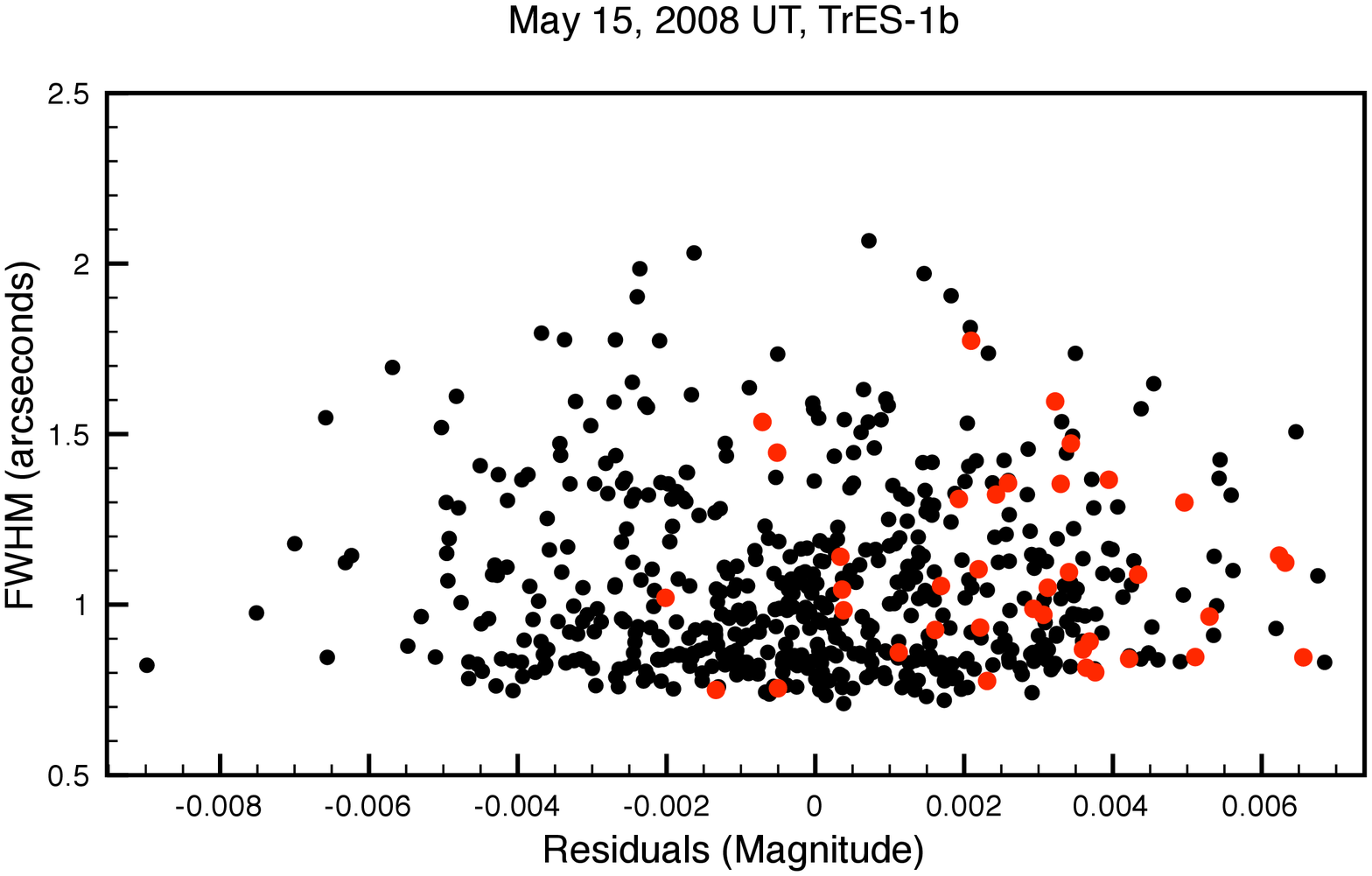}
\caption{
Scatter plot of the FWHM of the TrES-1 PSF vs. the unbinned residuals from the transit fit, May 12 on the left and May 15 on the right. Red points represent the anomaly photometry points. The correlation coefficient from Neter et al. (1982) between the points are quite low (-0.25 and 0.022 for May 12 and May 15 respectively). Considering only the red data points, the coefficients are similarly low, -0.212 for May 12 and -0.049 for May 15. Therefore we conclude that systematic effects associated with the atmosphere are not significantly affecting our photometry during (or even outside) of the anomalies.
}
\label{fwhm}
\end{figure}

\begin{figure}
\epsscale{1.0}
\plottwo{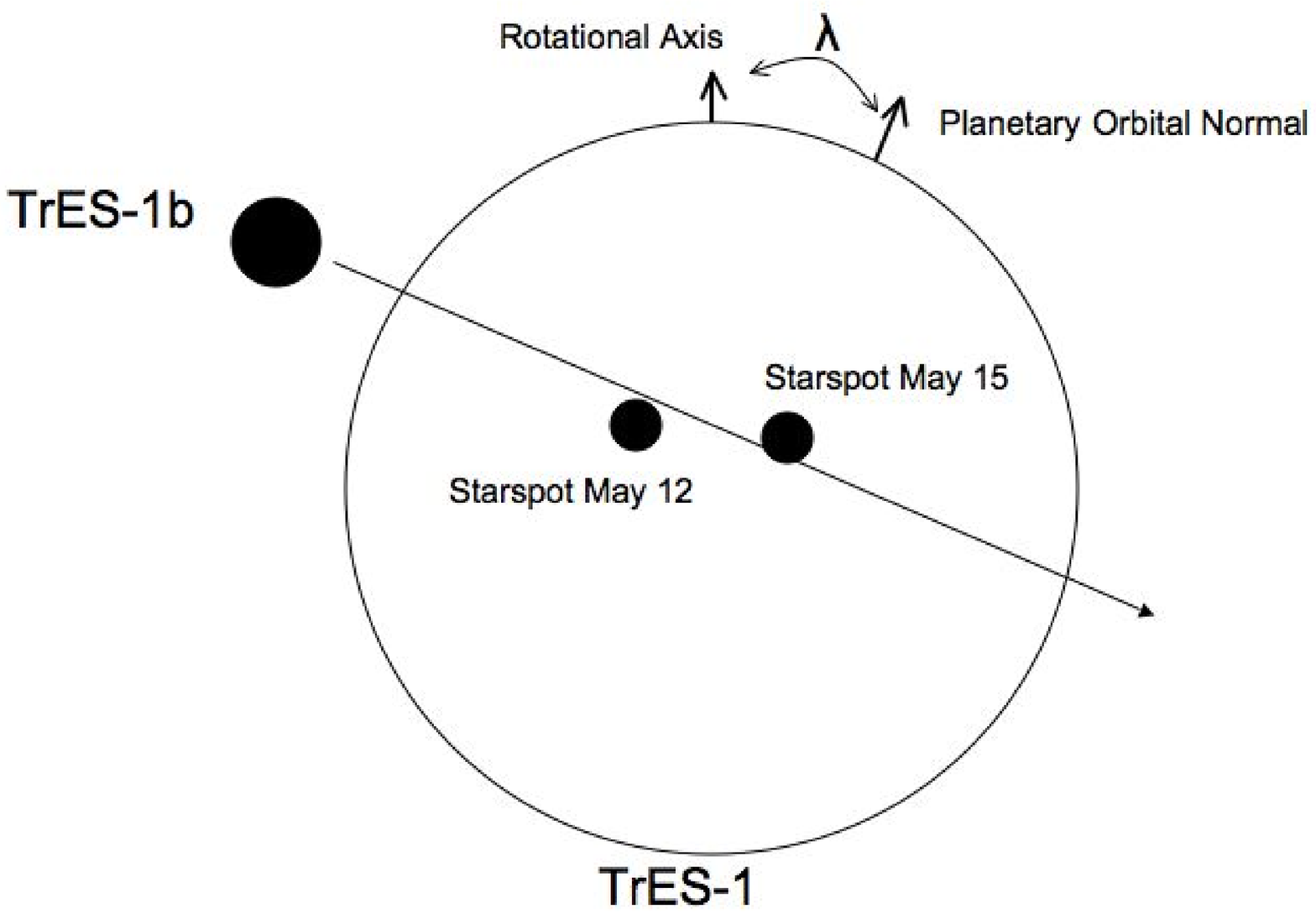}{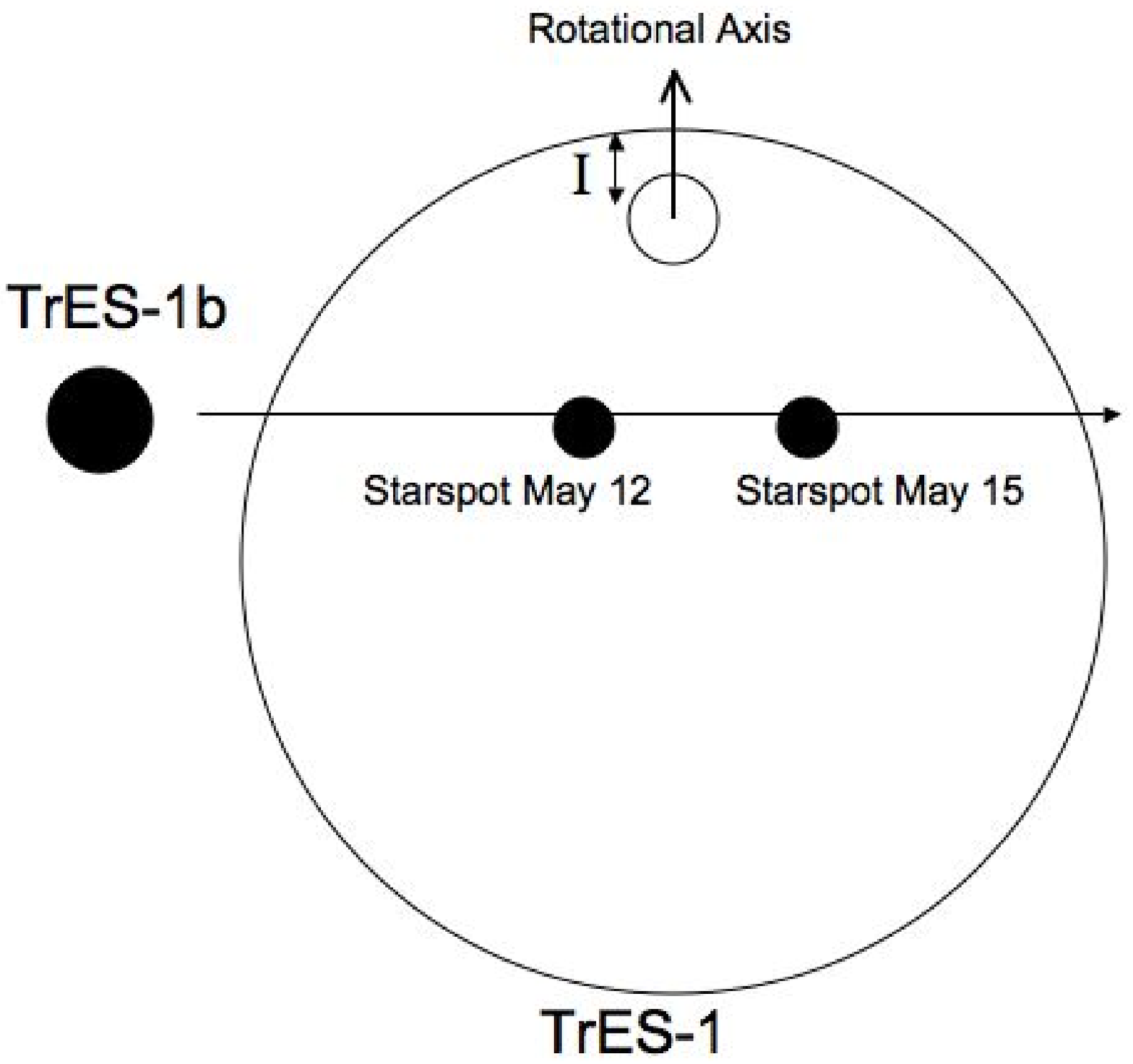}
\caption{
\textbf{Left}: Geometry for the nonzero $\lambda$ and $I=0^{\circ}$ case. While the time stamp of each starspot occultation remains the same, the longitudinal distance between them is shortened by a factor of cos($\lambda$) from the $\lambda = I = 0$ case (See Figure \ref{fig1}). \textbf{Right}: Geometry for the $\lambda = 0$ and nonzero $I$ case. Because the rotation axis of the star is tilted out of plane of the sky, the subtended longitudinal angle between starspot occultations is larger by a factor of $1$/cos($I$) than that inferred from the time stamp alone and the $\lambda = I = 0$ geometry. }
\label{angles}
\end{figure}

In order to understand the effect of red-noise at the frequency of interest to these anomalies, our transits are binned with a discrete bin to the red-noise frequency of interest (5 min periods, 15 minutes at the ingress of the May 12 transit due to the high air-mass of that observation) and then we investigate if these anomalies are significant by fitting a Gaussian to the residuals left from the transit fit. These Gaussian fits are shown in Figure \ref{gauss}. The peak of the spots are detected at $3.2 \sigma$ on May 12 and $2.9 \sigma$ on May 15. Other significant deviations from the transit (see Figure \ref{transits}) are shown as well. These other deviations are significant at the 1.3-1.7 $\sigma$ level. A summary of these deviations and their significance is shown in Table \ref{deviate}. We note that if we make the \textit{ad-hoc} assumption that the spot signal is real and refit a Gaussian to the residuals without the "spot" points, the significance of the spots increase to $3.4 \sigma$ and $3.1 \sigma$ for May 12 and May 15 respectively. 

We note that the significance of the signal we attribute to starspot occultation is $\geq 1.5 \sigma$ higher than other peak deviations which we associate with systematic red noise. Therefore, we estimate that each of these events have a $\sim 9\%$ chance each of being a very strong red noise peak. While this may seem like low significance, it is quite interesting to note that the May 15 spot falls within 30 minutes of the expected "position" for a real spot, given the May 12 position and a $P_{obs} = 33.2$ day rotation period with $\lambda = I = 30^{\circ}$. While we are at the limit of what can be achieved with a 1.6 m telescope in $\sim 10$ minutes, we feel that there is a significant ($>90\%$) chance this pair are the same spot. Yet it is impossible to be $100\%$ positive and it is also impossible to re-observe these events. However, the appearance of 3 new brightening anomalies in addition to that observed by HST strongly suggests large spots are not rare on the surface of TrES-1. 

\section{Conclusion}
During observations of two consecutive transits of TrES-1b, we found a brightening anomaly which we attributed to the planet crossing in front of a $\geq 6 R_{\earth}$ radius starspot. By observing the timing of TrES-1b eclipsing this spot during each transit, we can constrain the rotation rate of the star. Assuming that these are the same spot and that the spin axis of the star and orbital axis of the planet are aligned and in the plane of the sky suggests a stellar rotational period of $40.2 \pm 0.1$ days. Using the $\lambda = 30 \pm 21^{\circ}$ inclination of the stellar spin axis with respect to the planetary orbital axis of Narita et al. (2007) there is much more uncertainty and we are we are able to constrain the rotational period of the star to $40.2^{+22.9}_{-14.6}$ days, which is consistent with the previously observed $P_{obs} = 33.2^{+22.3}_{-14.3}$ day period. We note that this is a tentative detection of consecutive star spot occultations where the events associated with the spot are $\sim 1.5 \sigma$ higher in significance than red noise peaks in the light curve. We estimate that there is a $\sim 9\%$ chance that each event is red noise and a $\sim 1\%$ chance that both events are red noise. In the future, this technique for measuring stellar rotations can be applied to any transiting system in which significant sized star spots are eclipsed by the planet's transiting path, then rotation rate can be constrained. If the inclination of the stellar spin axis of the star is accurately known relative to the plane of the sky, this method could be a very accurate way to measure the rotational period of a star.

We have ruled out the possibility of these anomalies being a double planet occultation. If both starspot events were to be attributed to a double planet occultation, we find that the planet would have to have an orbital distance of $\sim 745$ AU and a period of $\sim 21000$ years. This is incompatible with a similar occultation observed by Rabus et al. (2009) in 2004, and therefore we conclude that the starspot explanation is the most likely to describe all events and there is no second outer transiting planet, TrES-1c, in the system. 

\section{Acknowledgments}
The authors would like to acknowledge the help of Trevor Olson and Louis Scuderi in collecting the data used in this work. We would like to thank the anonymous referee whose comments lead to a much improved paper. We would also like to thank the Arizona NASA Space Grant program for funding this work. LMC is supported by a NSF Career award and the NASA Origins program.

\clearpage

\begin{table}
\begin{center}
\caption{Parameters of the TrES-1 System} \label{tc}
\begin{tabular}{crrr}
\tableline\tableline
Parameter & Value & Ref. \\ 
\tableline
$P_{\mbox{orbit}}$ & $3.030065 \pm 8$ x $10^{-6}$ days & Alonso et al. (2004)\\
$a$ & $0.0393 \pm 0.0011$ AU & Alonso et al. (2004)\\ 
$M_{p}$ & $0.75 \pm 0.07 M_{Jup}$ & Alonso et al. (2004) \\ 
$R_{p}$ & $1.08^{+0.18}_{-0.04} R_{Jup}$ & Alonso et al. (2004) \\ 
$T_{\mbox{transit}}$ & 0.115 days & Alonso et al. (2004) \\ 
Latitude & $18^{\circ}$ & Alonso et al. (2004) \\ 
$R_{p}/R_{*}$ & $0.130^{+0.009}_{-0.003}$ & Alonso et al. (2004)\\ 
$P_{obs}$ & $38.4^{+12.7}_{-8.4}$ days$^{a}$ & Laughlin et al. (2005) \\ 
$P_{obs}$ & $38.4^{+12.7}_{-8.4}$ days$^{b}$ & Narita et al. (2007), Laughlin et al. (2005)\\ 
$P_{obs}$ & $33.2^{+22.3}_{-14.3}$ days$^{c}$ & Narita et al. (2007), Laughlin et al. (2005)\\ 
$P_{obs}$ & $33.2^{+22.3}_{-14.3}$ days$^{d}$ & Narita et al. (2007), Laughlin et al. (2005)\\ 
\tableline
$T_{\mbox{anomaly,1}}$ & 2454598.80273 JD & This work\\ 
$T_{\mbox{anomaly,2}}$ & 2454601.85759 JD & This work\\ 
$\theta_{1}$ & $0.2783$ rad\tablenotemark{a} & This work \\ 
$\theta_{2}$ & $0.1990$ rad\tablenotemark{a} & This work \\ 
$T_{c}$ April 14, 2007 & 2454204.91022 JD & This work\\
$T_{c}$ May 12, 2008 & 2454598.81648 JD & This work\\ 
$T_{c}$ May 15, 2008 &2454601.84757 JD & This work\\ 
$P_{\mbox{rot}}$ & $40.2 \pm 0.1$ days\tablenotemark{a} & This work \\ 
$P_{\mbox{rot}}$ & $46.4^{+23.6}_{-0.5}$ days\tablenotemark{b} & This work \\ 
$P_{\mbox{rot}}$ & $34.8^{+4.9}_{-9.5}$ days\tablenotemark{c} & This work \\ 
$P_{\mbox{rot}}$ & $40.2^{+22.9}_{-14.6}$ days\tablenotemark{d} & This work \\
\tableline
\end{tabular}

\tablenotetext{a}{assuming $\lambda = 0 \pm 0^{\circ}$, $I = 0 \pm 0^{\circ}$, $i_{*} = 90^{\circ}$}
\tablenotetext{b}{assuming $\lambda = 30 \pm 21^{\circ}$, $I = 0 \pm 0^{\circ}$, $i_{*} = 90^{\circ}$}
\tablenotetext{c}{assuming $\lambda = 0 \pm 0^{\circ}$, $I = 30 \pm 21^{\circ}$, $i_{*} = 60 \pm 21^{\circ}$}
\tablenotetext{d}{assuming $\lambda = 30 \pm 21^{\circ}$, $I = 30 \pm 21^{\circ}$, $i_{*} = 60 \pm 21^{\circ}$}

\end{center}
\end{table}

\clearpage

\begin{table}
\begin{center}
\caption{Detections of Brightening Anomalies During Transits of TrES-1b}\label{table}
\begin{tabular}{crrrrrr}
\tableline\tableline
Date & Height & Duration & Filter & Author \\ 
\tableline
(UT) & (mmag) & (min) &     &     \\ 
November 19, 2004 & $2.7 \pm 0.2$ & $\sim$10 & 450 - 900 nm & Rabus et al. (2009) \\ 
April 14, 2007 & $2.3 \pm 1.0$ & $\sim$11.5 & R & This work \\ 
May 12, 2008 & $5.4 \pm 1.7 $& $\sim$20.7 & R & This work \\ 
May 15, 2008 & $5.4 \pm 1.4$& $\sim$20.7 & R & This work \\
\tableline
\end{tabular}
\end{center}
\end{table}

\clearpage

\begin{table}
\begin{center}
\caption{List of Significant Deviations From Transit due to Correlated Noise and Starspot Occultation}\label{deviate}
\begin{tabular}{crrrr}
\tableline\tableline
Date & Number (See Figure \ref{transits}) & Positive/Negative Peak & $\sigma$ \\ 
\tableline
May 12 & Spot & Positive & 3.2 \\ 
May 15 & Spot & Positive & 2.9 \\ 
May 12 & Noise 1 & Negative & 1.7 \\ 
May 15 & Noise 2 & Negative & 1.5 \\ 
May 15 & Noise 3\tablenotemark{a} & Positive & 1.5 \\ 
May 15 & Noise 4\tablenotemark{b} & Positive & 1.3 \\ 
May 15 & Noise 5 & Negative & 1.7 \\
\tableline
\end{tabular}
\tablenotetext{a}{This \#3 noise peak could be due to another spot rotating onto the surface.}
\tablenotetext{b}{This \#4 noise peak could be due to a slightly poor egress fit with the $T_{transit}$ time of Alonso et al. (2004).}
\end{center}
\end{table}

\clearpage

\clearpage

\clearpage

\clearpage

\clearpage

\clearpage

\end{document}